# On Acoustic Dynamic Equation and Constitutive Relations in Thermo-piezoelectric and Dissipative Media

Yinqiu Zhou[1,2,3]††[*], Xiuming Wang[1,2,3]††, and Yuyu Dai[1,2,3]

[1]*State Key Laboratory of Acoustics, Institute of Acoustics, Chinese Academy of Sciences, Beijing 100190, China*
[2]*School of Physics Sciences, University of Chinese Academy of Sciences, Beijing 100149, China*
[3]*Beijing Engineering Research Center for Drilling and Exploration, Institute of Acoustics, Chinese Academy of Sciences, Beijing 100190, China*

In this work, a methodology is proposed for formulating acoustic dynamic equation with its constitutive relations in thermo-piezoelectric and dissipative media from the first principle of energy conservation and associated state function definitions, rather than using Hamilton's and/or variational principles. The main framework is based on energy conservation for piezoelectric medium. The results obtained from Hamilton's and/or virtual work principles and those from our energy conservation based methodology are in good agreement with each other. However, our formulations based on energy conservation are much easier to understand than using Hamilton's variational principle, where Lagrangian function densities and variational principles, associated with seeking extremum of a functional have to be introduced. What is more is that, the energy conservation based framework in formulating dynamic equation of motion is able to handle dissipation problems, which is usually behind the scope of Hamilton's principle. In our case, the mechanical, electric and thermal phenomena firstly are taken into account, and then the medium with dissipation is included. The advantage of our formulations is that only the conservation of kinetic energy, potential energy, heat and work of a closed continuum is taken into account, and all of the dynamic equations and their constitutive relations can be derived if the formal state function of the continuum are defined. Our formulations on acoustic dynamic equation and the associated constitutive relations of the medium will also pave an alternative way in computational dynamic modelling based on weak formulations, such as finite element and spectral element methods. In addition, our methodology may be extended to other dynamic equation formulations, such as those in electrodynamics, fluid mechanics and quantum mechanics. This is especially true for tackling the problems with multi-physical field interactions and coupling.

## 1 Introduction

Acoustic wave propagation in thermo-piezoelectric media governed by dynamic equations has been tackled for more than 80 years. Establishment of the wave equation of motion and associated constitutive relations have been investigated thoroughly based on either Newton's second Law, or Hamilton's principles. For more complicated situations, variational principles or generalized Hamilton's principles have been successfully introduced [Tiersten,1967; EerNisse,1967; Mindlin,1972]. More discussions have been seen on how to construct Lagrangian density function in order to use the Hamilton's principle to derive acoustic wave equation and its corresponding boundary value conditions [Holland,1968; Lazerson,1975; Gubenkov,1978].

As is known, Hamilton mechanics is so powerful that it converts the state of the system into generalized coordinates $q_i$, and changes in those coordinates, $dq_i/dt$. These coordinates represent a simplified phase space for the system. Since Lagrangian function is a scalar representation of a physical system's position in phase space, and changes in the Lagrangian reflect the movement of the system in phase space. Because it's a single number, this principle makes the related equations far simpler [Goldstein, 2011].

Unfortunately, Hamilton's principle, could not be directly applied to those that the Lagrangian function could not be easily formulated because of the external forces or non-conservative system. The generalized Hamilton's principle has been developed to characterize wave propagation. However, how to take into account mechanical, piezoelectric, electromagnetic and thermal effects has been always questionable, such as shown in [Tiersten,1967; Zhang,1985; Altay,2005; Preumont,2006] Even for a conservative system, the applicability of Hamilton's principle has been questioned to some extent [Martin,2017] in predicting the natural trail of a body's movement. What is more, because of the complexity of wave propagation in thermo-piezoelectric dissipative media, more general methods, such as variational principle has been used for describing various energy interactions, such as elastic, electromagnetic, and thermal couplings [Tiersten,1969; Mindlin,1972; Auld,1973]. The variational principle, itself is based on purely mathematic knowledge, i.e., to seek extremum value of a functional belongs to functional analyses [Gelfand and Fomin, 1963]. The problem is that how to construct the functional that converts physical principle into problem of calculus of variations [Pramila,1992]. So it turns out that one still has to find a pertinent functional based on Hamilton's or generalized Hamilton's principles. What is more is that Hamilton's principle has limited conditions that was discussed by several authors, such as it has a limit in which one needs to take into account the boundary conditions and so on [Courant,1953; Morse,1953; Rosen,1954; Tiersten,1968; Dökmecí 1980, Luo and Li, 2010].

In this work, by taking into the prons and cons of the formulations of acoustic wave equation and so on, we reconsider the establishment of acoustic wave equation of motion and the related constitutive relations for thermo-piezoelectric media from the first principle, i.e., the first law of thermodynamics, rather than using Hamilton's principle and/or various variational principles. We even don't touch Newton's second law of motion to construct these equations. The advantage by doing so is that only the conservation of kinetic energy, internal energy, heat, and work of a closed continuum system is taken into account, and all of the wave equations of motion and their constitutive relations could be derived, if the state function of the continuum system are formally defined.

## 2 Acoustic Wave Equation and Constitutive Equations

### 2.1 From energy conservation principle to acoustic wave equation of motion

Based on the principle of conservation of energy for a thermo-piezoelectric medium, the increase of the total energy $dE$ (including kinetic energy and internal energy) in any element volume $V$ bounded by a surface $S$ with unit outward normal $n_i, i = 1, 2, 3,$ is equal to the external work $dW$ done on the volume of the medium plus the thermal energy $dQ$ delivered to the medium [Mindlin,1961]. The external work includes the work done by body forces, surface tractions and the electric forces due to the surface electric charge. Suppose $f_i$ is the body force; $t_{ij}$, the contact force; $\rho$ is the mass density; $u_i$ is the displacement of a point; $U$ is the internal energy; $\varphi$ is the electric potential; $D_i$ is electric displacement; $\theta$ is temperature and $\sigma$ is entropy. The law of energy conservation is written as

$$dE = dQ + dW, \text{or}$$
$$dE/dt = dQ/dt + dW/dt. \quad (1)$$

Equation (1) states that the rate of the total energy with respect to time is equal to the rate of the work done on the volume of the medium, with respect to time, plus the rate of heat delivered to the medium with respect time.

According to the definition of kinetic energy density function, $K = \rho \dot{u}_i \dot{u}_i / 2$, and internal energy one, denoted as $U$, the state function of various physical variables. It is noted that the internal energy is assumed that it is not a function of time $t$ explicitly. Therefore, the energy conservation law, by using Einstein's convention is written as

$$\partial \left( \iiint_V (\rho \dot{u}_i \dot{u}_i / 2 + U) dV \right) / \partial t$$
$$= \iiint_V (f_i \dot{u}_i) dV + \iint_S (t_i \dot{u}_i - n_i \varphi \dot{D}_i) dS + \iiint_V \theta \dot{\sigma} dV, \quad (2)$$

where the first term on the left hand side of the above equation represents the total energy rate with respect to time $t$, and the first, second and the third terms on the right hand side of the equation, are the external work rate with respect to time $t$ done by body forces, surface tractions, and electric forces due to the surface electric charge, respectively, while the last term stands for the thermal energy rate with respect to time injected into the volume $V$. Since the traction force is written as $t_i = n_j \tau_{ji}$, Equation (2) can be rearranged

$$\iiint_V (\rho \dot{u}_i \ddot{u}_i + \dot{U}) dV = \iiint_V (f_i \dot{u}_i) dV$$
$$+ \iint_S (n_j \tau_{ji} \dot{u}_i - n_i \varphi \dot{D}_i) dS + \iiint_V \theta \dot{\sigma} dV. \quad (3)$$

By using Gauss's divergence theorem, the above equation is converted into

$$\iiint_V (\rho \dot{u}_i \ddot{u}_i + \dot{U}) dV = \iiint_V (f_i \dot{u}_i) dV$$
$$+ \iiint_V [(\tau_{ji} \dot{u}_i)_{,j} - (\varphi \dot{D}_i)_{,i}] dV + \iiint_V \theta \dot{\sigma} dV. \quad (4)$$

Therefore,

$$\iiint_V (\rho \dot{u}_i \ddot{u}_i + \dot{U}) dV = \iiint_V (f_i \dot{u}_i) dV$$
$$+ \iiint_V (\tau_{ji,j} \dot{u}_i + \tau_{ji} \dot{u}_{i,j} - \varphi_{,i} \dot{D}_i - \varphi \dot{D}_{i,i} + \theta \dot{\sigma}) dV. \quad (5)$$

Rearranging the above equation, yields

$$\iiint_V (\rho \ddot{u}_i - f_i - \tau_{ji,j}) \dot{u}_i dV$$
$$+ \iiint_V (\dot{U} - \tau_{ji} \dot{u}_{i,j} + \varphi_{,i} \dot{D}_i + \varphi \dot{D}_{i,i} - \theta \dot{\sigma}) dV = 0. \quad (6)$$

Naturally, the second term is zero because it gives the definition of state function $U$ of the system. Therefore, Equation (6) becomes,

$$\iiint_V (\rho \ddot{u}_i - f_i - \tau_{ji,j}) \dot{u}_i dV = 0.$$

Since the above equation holds for any given volume $V$, $\dot{u}_i$ does not always equal 0, it is easy to shown that

$$\rho \ddot{u}_i - f_i - \tau_{ji,j} = 0, \quad (7)$$

which is the acoustic wave equation of motion. From Equation (6), as the state function definition, we also obtain

$$\dot{U} = \tau_{ji} \dot{u}_{i,j} + \varphi_{,i} \dot{D}_i + \varphi \dot{D}_{i,i} - \theta \dot{\sigma}, \quad (8)$$

which eventually leads to the constitutive equation of the given thermo-piezoelectric medium of interest. Since for electrostatic state, there is no free charge in the volume of interest, we have

$$\dot{D}_{i,i} = 0, \qquad (9)$$
$$E_i = -\varphi_{,i}. \qquad (10)$$

Since we don't take into account rotation motion, the angular moment is conserved. Therefore, $\tau_{ij} = \tau_{ji}$. The definition of strain is $S_{ij} = (u_{i,j} + u_{j,i})/2$. Hence we have the expression of $\dot{U}$ written as

$$\dot{U} = \tau_{ji}\dot{S}_{ij} + E_i\dot{D}_i + \theta\dot{\sigma} \qquad (11)$$

By internal energy definition, we may give general constitutive relations of the medium of interest. However, there are several descriptions for defining the state function with various group co-variables. Any kind of group selections does not affect the discussion results, and more discussions can be seen in [Auld,1973]. Therefore, here we define $U = U(S_{ij}, D_i, \sigma)$. Hence, we have

$$dU = (\partial U/\partial S_{ij})dS_{ij} + (\partial U/\partial D_i)dD_i \\ + (\partial U/\partial \sigma)d\sigma \qquad (12)$$

If we denote $\tau_{ij} = \partial U/\partial S_{ij}, E_i = \partial U/\partial D_i, \theta = \partial U/\partial \sigma$, then Equation (12) is rewritten as

$$dU = \tau_{ij}dS_{ij} + E_i dD_i + \theta d\sigma, \qquad (13)$$

Equation (13) is also called the first law of thermodynamics for a piezoelectric medium if there is no dynamic behavior in the given volume $V$, i.e., $\dot{u}_i = 0$.

## 2.2 Representation of piezoelectric constitutive relations

Let us define electric enthalpy $H$ by $H = U - E_i D_i$, then differentiating it with respect to time, we obtain

$$\dot{H} = \dot{U} - E_i\dot{D}_i - \dot{E}_i D_i \qquad (14)$$

Substituting Equation (14) into Equation (11), yields

$$\dot{H} = \tau_{ij}\dot{S}_{ij} + \theta\dot{\sigma} - D_i\dot{E}_i \qquad (15)$$

As is mentioned early, selection of different group of co-variables will not have an effect on the discussion results, we consider the electric enthalpy as a state density function of $S_{ij}, E_i, \sigma$, i.e., $H = H(S_{ij}, E_i, \sigma)$. Then, differentiating $H$ with respect to time, yields

$$\dot{H} = (\partial H/\partial S_{ij})_{E,\sigma}\dot{S}_{ij} + (\partial H/\partial E_i)_{S,\sigma}\dot{E}_i \\ + (\partial H/\partial \sigma)_{S,E}\dot{\sigma} \qquad (16)$$

Subtracting Equation (16) from Equation (15), yields

$$[\tau_{ij} - (\partial H/\partial S_{ij})_{E,\sigma}]\dot{S}_{ij} - [D_i + (\partial H/\partial E_i)_{S,\sigma}]\dot{E}_i \\ + [\theta - (\partial H/\partial \sigma)_{S,E}]\dot{\sigma} = 0. \qquad (17)$$

Equation (17) must hold for arbitrary $\dot{S}_{ij}, \dot{E}_i, \dot{\sigma}$. Therefore,

$$\tau_{ij} = (\partial H/\partial S_{ij} + \partial H/\partial S_{ji})/2 = (\partial H/\partial S_{ij})_{E,\sigma},$$
$$D_i = -(\partial H/\partial E_i)_{S,\sigma}, \theta = (\partial H/\partial \sigma)_{S,E}. \qquad (18)$$

We expand the above three formulas separately and only retain the linear terms. Hence,

$$\begin{cases} d\tau_{ij} = (\partial \tau_{ij}/\partial S_{kl})_{E,\sigma}dS_{kl} + (\partial \tau_{ij}/\partial E_k)_{S,\sigma}dE_k \\ \quad + (\partial \tau_{ij}/\partial \sigma)_{S,E}d\sigma, \\ dD_i = (\partial D_i/\partial S_{kl})_{E,\sigma}dS_{kl} + (\partial D_i/\partial E_k)_{S,\sigma}dE_k \\ \quad + (\partial D_i/\partial \sigma)_{S,E}d\sigma, \\ d\theta = (\partial \theta/\partial S_{kl})_{E,\sigma}dS_{kl} + (\partial \theta/\partial E_k)_{S,\sigma}dE_k \\ \quad + (\partial \theta/\partial \sigma)_{S,E}d\sigma. \end{cases} \qquad (19)$$

In these equations, all the constants are defined as below:

$(\partial \tau_{ij}/\partial S_{kl})_{E,\sigma} = c_{ijkl}^E$ is the Elastic constant component under isentropic process and constant electric field;

$(\partial \tau_{ij}/\partial E_k)_{S,\sigma} = -e_{ijk}$ is the Piezoelectric constant component under adiabatic process and constant strain;

$(\partial \tau_{ij}/\partial \sigma)_{S,E}d\sigma = (\partial \tau_{ij}/\partial \sigma)_{S,E}dQ/\theta = -\gamma_{ij}^E dQ$, where $\gamma_{ij}^E$ is the piezocaloric coefficient under constant electric field; $(\partial D_i/\partial S_{kl})_{E,\sigma} = e_{ikl}$ is the Piezoelectric constant component under adiabatic process and constant electric field; $(\partial D_i/\partial E_k)_{S,\sigma} = \varepsilon_{ik}^S$ is the Dielectric constant component under adiabatic process and constant strain;

$(\partial D_i/\partial \sigma)_{S,E}d\sigma = (\partial D_i/\partial \sigma)_{S,E}dQ/\theta = -q_i^S dQ$, where $q_i^S$ is Electric thermal constant under constant strain;

$(\partial \theta/\partial S_{kl})_{E,\sigma}/\theta = (\partial \tau_{kl}/\partial \sigma)_{S,E}/\theta = -\gamma_{kl}^E$, where $\gamma_{kl}^E$ is the Thermal compression coefficient under constant electric field;

$(\partial \theta/\partial E_k)_{S,\sigma}/\theta = -(\partial D_k/\partial \sigma)_{S,E}/\theta = -q_k^S$, where $q_k^S$ is the Thermal electric constant under constant strain;

$(\partial \theta/\partial \sigma)_{S,E}d\sigma = (\partial \theta/\partial \sigma)_{S,E}dQ/\theta = (1/\rho C_V^E)dQ$, where $\rho$ is density and $C_V^E$ is Specific heat of capacitance under constant electric field.

We substitute all of the above constants into Equation (19), $\tau_{ij}, D_i, \theta$ are represented as

$$\begin{cases} \tau_{ij} = c_{ijkl}^E S_{kl} - e_{ijk}E_k - \gamma_{ij}^E dQ, \\ D_i = e_{ikl}S_{kl} + \varepsilon_{ik}^S E_k - q_i^S dQ, \\ d\theta = -\theta\gamma_{kl}^E S_{kl} - \theta q_k^S E_k + (1/\rho C_V^E)dQ. \end{cases} \qquad (20)$$

For the adiabatic process, $dQ = 0$, then we obtain the linear piezoelectric constitutive equations

$$\begin{cases} \tau_{ij} = c_{ijkl}^E S_{kl} - e_{ijk}E_k, \\ D_i = e_{ikl}S_{kl} + \varepsilon_{ik}E_k. \end{cases} \qquad (21)$$

Equation (21) are general piezoelectric equation, and the expression of temperature increase is

$$d\theta = -\theta\gamma_{kl}^E S_{kl} - \theta q_k^E E_k. \qquad (22)$$

Based only on the law of energy conservation, we have established dynamic equation, and all of the constitutive equations. It is noted that, different from the existing procedure, we never touched either generalized Hamilton's variational principles, least action principles, or Newton's second law of motion. In order to validate our results, the followings are shown to derive acoustic wave dynamic equations and associated constitutive equations based on generalized Hamilton's variational principle.

**2.3 Applying Hamilton's principle**

For an open continuum, the Lagrangian formulation for a holonomic potential system or any given enclosed volume in a continuous elastic medium is represented as [Achenbach, 1975; Liu, 2020]

$$d(\partial L/\partial \dot{q}_i)/dt + (\partial L/\partial q_{i,j})_{,j} - \partial L/\partial q_i = 0, \quad (23)$$

where $L$ is Lagrangian density function, and $q_i, i=1,2,3$, the generalized displacement component. It seems that, once the Lagrangian density function is identified, the wave equation of motion will be directly derived from the generalized Lagrangian equation in Equation (23). However, one may not directly use this equation because in thermo-piezoelectric medium, there are external forces that may do work on the system of interest. Therefore, we need to go from general sense by using D'Alembert's principle which is based on Newton's second law of motion, to establish generalized Hamilton's principle, or the variational principle in deriving acoustic dynamic equations and related boundary conditions. When the system is not conservative, the principle can be generalized by calculating the virtual work $\delta W$ done by the non-conservative forces in a displacement consistent with the constraints [Tiersten,1967].

$$\int_{t_0}^{t} \delta L dt + \int_{t_0}^{t} \delta W dt = 0. \quad (24)$$

That is, the above so called generalized variable principle is established using the dynamic equation, or Newton's second law of motion, and it is natural that the derived dynamic equations must be held because the pre-existing condition of the second law of motion has been included, although Hamilton's principle itself does not need to introduce it. First, we identify kinetic and potential energy functions from its density function, i.e.,

$$T = \iiint_V (\rho \dot{u}_i \dot{u}_i / 2) dV, \text{ and} \quad (25a)$$

$$V = \iiint_V H(S_{ij}, E_i, \sigma) dV. \quad (25b)$$

Then, the Lagrangian function can be written as

$$L = \iiint_V (\rho \dot{u}_i \dot{u}_i / 2 - H) dV \quad (26)$$

On the other hand, for any given virtual displacement $\delta u_i$ and virtual electric potential $\delta \varphi$, the virtual work done mechanically and electrically is written as

$$\delta W = \iint_S (\bar{t}_i \delta u_i - \bar{\sigma} \delta \varphi) dS + \iiint_V (f_i \delta u_i) dV. \quad (27)$$

Therefore, Equation (24) can be reformed in detail as

$$\begin{aligned}
&\delta \int_{t_0}^{t} dt \iiint_V (\rho \dot{u}_i \dot{u}_i / 2) dV \\
&-\delta \int_{t_0}^{t} dt \iiint_V H(S_{ij}, E_i) dV \\
&+\int_{t_0}^{t} dt \iint_S (\bar{t}_i \delta u_i - \bar{\sigma} \delta \varphi) dS \\
&+\int_{t_0}^{t} dt \iiint_V (f_i \delta u_i) dV = 0,
\end{aligned} \quad (28)$$

where $\bar{t}_k$, $\bar{\sigma}$ and all the variations vanish at $t_0$ and $t$. The first term in the above expression can be written as,

$$\begin{aligned}
&\delta \int_{t_0}^{t} dt \iiint_V (\rho \dot{u}_i \dot{u}_i / 2) dV \\
&= \int_{t_0}^{t} dt \iiint_V [\partial(\rho \dot{u}_i \delta u_i)/\partial t - \rho \ddot{u}_i \delta u_i] dV \\
&= \iiint_V (\rho \dot{u}_i \delta u_i |_{t_0}^{t} dV) - \int_{t_0}^{t} dt \iiint_V \rho \ddot{u}_i \delta u_i dV \\
&= -\int_{t_0}^{t} dt \iiint_V \rho \ddot{u}_i \delta u_i dV,
\end{aligned} \quad (29)$$

where the boundary conditions with zero virtual displacements have been used. By using the electric enthalpy $H$, we obtain that

$$\dot{H} = \tau_{ij} \dot{S}_{ij} + \theta \dot{\sigma} - D_i \dot{E}_i \quad (30)$$

The second term in Equation (28) can also be written as

$$\begin{aligned}
&\delta \int_{t_0}^{t} dt \iiint_V H(S_{ij}, E_i, \sigma) dV = \\
&= \int_{t_0}^{t} dt \iint_S (n_k \tau_{kl} \delta u_l + n_k D_k \delta \varphi) dS \\
&-\int_{t_0}^{t} dt \iiint_V (\tau_{kl,k} \delta u_l + D_{k,k} \delta \varphi - \theta \delta \sigma) dV
\end{aligned} \quad (31)$$

Substituting Equations (29) and (31) into Equation (28), we find

$$\begin{aligned}
&\int_{t_0}^{t} dt [\iiint_V (\tau_{kl,k} - \rho \ddot{u}_l + f_l) \delta u_l dV + \iiint_V D_{k,k} \delta \varphi dV \\
&-\iiint_V \theta \delta \sigma dV + \iint_S (\bar{t}_l - n_k \tau_{kl}) \delta u_l dS \\
&-\iint_S (n_k D_k + \bar{\sigma}) \delta \varphi dS] = 0.
\end{aligned} \quad (32)$$

Hence the above equation shows that,

$$\rho \ddot{u}_l - \tau_{kl,k} - f_l = 0, \quad (33)$$

which is acoustic wave equation of motion, and

$$D_{k,k} = 0. \quad (34)$$

And also, since $\delta \sigma$ is not always 0, Equation (32) shows that

$$\theta = T - T_0 = 0. \quad (35)$$

Following the previous analysis, on the surface S, we obtain $\bar{t}_l - n_k \tau_{kl} = 0$ or $u_l$ is given, and $n_k D_k + \bar{\sigma} = 0$ or $\varphi$ is given.

Now we compare the two results in Equations (7) and (11), given by using energy conservation law and those given in Equations (30) and (33) by using generalized Hamilton's principle. Both of them whether its dynamic equations or the associated constitutive equations are perfectly agreed with each other. However, the procedure based on energy conservation law is much intuitive and simple because one does not use general coordinates and variational concepts. Further analysis shows that this weak form provided with the conservation law give a simple way for weak form representation that will be used in finite element method and so on.

## 3 Consideration of Dissipation Energy

**3.1 Energy conservation principle with dissipative forces**
If we consider dissipation energy, the work of unit volume done by a dissipative volume-like force $-b\dot{u}_i$ with a displacement $\dot{u}_i dt$ is written as

$$dW_D = -\iiint_V (b\dot{u}_i\dot{u}_i)dVdt, \quad (36)$$

where $b$ is friction coefficient. The energy conservation equation in a dissipative medium of $V$ can be written as

$$\partial\left(\iiint_V (\rho\dot{u}_i\dot{u}_i/2 + U)dV\right)/\partial t = $$
$$\iiint_V (f_i\dot{u}_i)dV + \iint_S (t_i\dot{u}_i - n_i\varphi\dot{D}_i)dS$$
$$+\iiint_V \theta\dot{\sigma}dV - \iiint_V (b\dot{u}_i\dot{u}_i)dV.$$

The above equation can be reformulated as

$$\iiint_V (\rho\ddot{u}_i - f_i - \tau_{ji,j} + b\dot{u}_i)\dot{u}_i dV$$
$$+\iiint_V (\dot{U} - \tau_{ji}\dot{u}_{i,j} + \varphi_{,i}\dot{D}_i + \varphi\dot{D}_{i,i} - \theta\dot{\sigma})dV = 0. \quad (37)$$

Following the procedure described in Section 2.1, for any given volume $V$, $\dot{u}_i$ does not always equal 0. Therefore, the first term on left hand side in Equation (37) requires that

$$\rho\ddot{u}_i - f_i - \tau_{ji,j} + b\dot{u}_i = 0, \quad (38)$$

which is acoustic wave equation of motion in dissipative thermo-piezoelectric continuous medium. Since $\dot{D}_{i,i} = 0$, $E_i = -\varphi_{,i}$, $\tau_{ij} = \tau_{ji}$, and $S_{ij} = (u_{i,j} + u_{j,i})/2$, using the state function definition, the total internal energy density function rate with respect to time is given in the following form

$$\dot{U} = \tau_{ji}\dot{u}_{i,j} - \varphi_{,i}\dot{D}_i + \theta\dot{\sigma} = \tau_{ij}\dot{S}_{ij} + E_i\dot{D}_i + \theta\dot{\sigma}, \quad (39)$$

which is the same as in Equation (11).

**3.2 Acoustic wave equation from generalized Hamilton's principle**
The generalized Hamilton's variational principle used here is similar to Equation (24). The only difference is that the dissipative forces are taken into account. Following the procedure described in Section 2.3, we first calculate $\delta L$ and $\delta W$, respectively. When dissipation is considered, the virtual work $\delta W_D$ done by the dissipative force is defined as

$$\delta W_D = -\iiint_V b\dot{u}_i\delta u_i dV, \text{ and}$$

$$\delta W_1 = \iint_S (\bar{t}_i\delta u_i - \bar{\sigma}\delta\varphi)dS + \iiint_V (f_i\delta u_i)dV. \quad (40)$$

Therefore, the virtual work $\delta W$ will change to

$$\delta W = \delta W_1 + \delta W_D = \iint_S (\bar{t}_i\delta u_i - \bar{\sigma}\delta\varphi)dS$$
$$+\iiint_V (f_i\delta u_i)dV - \iiint_V b\dot{u}_i\delta u_i dV. \quad (41)$$

By using the expression of $\dot{H}$ written as

$$\dot{H} = \tau_{ij}\dot{S}_{ij} + \theta\dot{\sigma} - D_i\dot{E}_i, \quad (42)$$

substituting $\delta W$ and $H$ into Equation (24) yields

$$\delta\int_{t_0}^{t} dt \iiint_V (\rho\dot{u}_i\dot{u}_i/2 - H)dV$$
$$+\int_{t_0}^{t} dt \iint_S (\bar{t}_k\delta u_k - \bar{\sigma}\delta\varphi)dS \quad (43)$$
$$+\int_{t_0}^{t} dt \iiint_V (f_i\delta u_i)dV + \int_{t_0}^{t} dt \iiint_V (b\dot{u}_i\delta u_i)dV = 0.$$

Following the procedure described in Section 2.3, we have

$$\rho\ddot{u}_l - \tau_{kl,k} - f_l + b\dot{u}_i = 0 \quad (44)$$

which is acoustic wave equation of motion, and

$$D_{k,k} = 0. \quad (45)$$

Also, since $\delta\sigma$ is not always 0, from Equation (32) we have

$$\theta = T - T_0 = 0. \quad (46)$$

On the surface S, we obtain $\bar{t}_l - n_k\tau_{kl} = 0$ or $u_l$ is given, and $n_k D_k + \bar{\sigma} = 0$ or $\varphi$ is given.

Now we make the comparison between the two results in Equations (38) and (39), given by using energy conservation law and those given in Equations (44) and (42), by generalized Hamilton's variational principle. We obtain the same acoustic wave equation of motion by using the same definition of internal energy density given by the first law of thermodynamics for a piezoelectric medium no matter which principle we use.

## 4 Conclusion and Discussions

We have established acoustic dynamic equation of motion and the constitutive relations for a thermo-piezoelectric dissipative medium from the first law of thermodynamics and energy conservation principle. By derivation, we deduced the acoustic wave equation of motion only based from energy conservation principle. We compared these results with those given by using generalized Hamilton's principle and/or virtual work principle. Both deduced results are in good agreement. The advantage of our equation formulations is that the conservation of kinetic energy, potential energy, heat, and work of a closed continuum system is taken into account, and all of the acoustic dynamic equations can be easily obtained if the generalized state function of a continuum are defined properly. Since energy conservation will take into account all of the energies and works in balance, our method is especially efficient in dealing with more complex dissipation problems in multi-physical field interaction and coupling.